\pdfoutput=1
\documentclass[isre, anonymous,figcaps]{informs3} 

\DoubleSpacedXI 

\usepackage{hyperref}       
\usepackage{url}            
\usepackage{booktabs}       
\usepackage{amsfonts}       
\usepackage{nicefrac}       
\usepackage{microtype}      
\usepackage{multirow}
\usepackage{graphicx}
\usepackage{float}
\usepackage{algorithm}
\usepackage{amsmath}
\usepackage{algorithmic}
\usepackage{amsfonts}
\usepackage{subfig}
\usepackage{lipsum}

\usepackage{natbib}
 \bibpunct[, ]{(}{)}{,}{a}{}{,}%
 \def\bibfont{\small}%
\newtheorem{definition}{Definition}

\begin{document}




\TITLE{Characterizing presence patterns and segmenting user locations from cell phone data}

\large
\ARTICLEAUTHORS{%
\AUTHOR{Yan Leng}
\AFF{MIT Media Lab\\  \EMAIL{yleng@mit.edu}, \URL{}}
\AUTHOR{Jinhua Zhao}
\AFF{MIT Department of Urban Studies and Planning\\ \EMAIL{jinhua@mit.edu}, \URL{}}
\AUTHOR{Haris N. Koutsopoulos}
\AFF{Northeastern University Civil and Environmental Engineering\\ \EMAIL{ h.koutsopoulos@northeastern.edu}, \URL{}}
} 

\ABSTRACT{%
The dynamic monitoring of home and workplace distribution is a fundamental building block for improving location-based service systems in fast-developing cities around the world. Inferring these places is a challenging task; existing approaches rely on labor-intensive and untimely survey data or ad-hoc heuristic assignment rules based on the frequency of appearance at given locations. In this paper, we propose a novel method to infer the home, workplace, and third place based on an individual's spatial-temporal patterns inferred from Call Detail Records. Using the historical data of geo-temporal travel patterns for a panel of individuals, our method develops, for each person-location, the probability distribution that the person will appear in that location at a specific time of day. We apply eigen-decomposition to the matrix of historical geo-temporal data. Unsupervised machine learning techniques are then used to extract commonalities across locations for different groups of travelers, and make inferences, such as home and workplace. Testing the methodology on real-world data with known location labels shows that our method identifies home and workplace with significant accuracy, improving upon the most commonly used methods in the literature by 79\% and 34\%, respectively. 
The methodology proposed is computationally efficient and is highly scalable to other real-world applications with historical tracking data. It provides a basis to improve location-based services, such as mobile commerce, social events recommendations, and urban transit design. 
}%


\KEYWORDS{Human mobility; Origin-Destination flow; Travel demand management; Unsupervised learning}

\maketitle

%

\section{Introduction}

Understanding human mobility and urban dynamics are essential for transit operators, urban planners, and location-based service providers~\citep{hasija2020smart}. 
Extensive penetration and rapid development of mobile technologies make it possible for
service providers and policy-makers for dynamic and contextual decision-making~\citep{macedo2015context,zhang2012contextual,andrews2016mobile}.
Among different aspects of human mobility, inferring the home and workplaces are the essential building blocks for a wide range of applications in management information systems~\citep{hong2006understanding}. 

First, these locations and the corresponding commuting flows are crucial for location-based mobile targeting, since the context and locations are essential for mobile marketing and  targeting~\citep{banerjee2008mobile}. 
Second, commuting flows are the significant contributor to peak-hour traffic flow and hence determine the bottleneck of transit supply. 
Third, as most mobile phone data lacks sociodemographics due to privacy reason, accurate inference on home locations enables the mapping to the census data with rich information. 
Lastly, understanding the spread of individual daily activities, detecting emerging residential and commercial regions, help policymakers better monitor urban dynamics and spatial-temporal distribution of the population.

However, existing approaches for inferring home and workplace at the individual and aggregate levels are not accurate and flexible enough, which rely on simplistic assumptions. 
Traditionally, the means to infer home and workplace distribution is either through household travel surveys, which is labor-intensive and costly~\citep{toole2014path}. 
The emergence of ubiquitous digital data collection infrastructure (e.g., cell towers, WiFi hotspots, or blue-tooth beacons) provides the opportunity to monitor human mobility and urban dynamics at an unprecedented spatial scale and granularity \citep{gr2015, blondel2015}. 
The existing method makes simplistic assumptions based on presence patterns at home and workplaces using mobile phone data. 
However, these methods might be inaccurate due to the sparsity nature of most mobile phone data and the complexities in human behavior, which we show in the experiment section of this paper. 
Several existing studies have shown that human behavior, especially mobility behaviors, are highly regular and predictable \citep{gonzalez2008understanding, lu2013approaching, schneider2013unravelling}. 
In this paper, we utilize such regularities to infer home and workplace, which is a challenging but crucial task, especially in fast-developing cities. 


We propose a methodology to extract regular behavioral patterns at locations and infer home/workplace in urban spaces based on Call Detail Records. 
This method is applied to Call Detail Records (CDR), an opportunistic, large-scale, and longitudinal data source. The high penetration rate of cell phones makes CDR an appropriate proxy for travel behaviors, which generates new opportunities in understanding aggregate urban dynamics and individual mobility. 
We have witnessed efforts in both academia and industry in using CDR in transportation and urban planning. Let us name a few studies. 
\cite{phi2010} use this data to understand the land usage pattern, precisely, whether activity patterns at similar locations are similar (or not). The information can further be used to estimate the most probable activities associated with particular regions of a city. \cite{calabrese2014} extracted travel activity information and infer activity patterns by merging mobile phone data and travel diary surveys.

Existing methodologies solve this problem by either relying on labor-intensive and untimely data or using ad-hoc heuristic assignment rules based on the frequency of appearances at given locations from CDR. In this research, we apply eigen-decomposition and unsupervised machine learning tools to large-scale mobility data in a populated city in China to extract commonalities in behavioral patterns across locations for different groups of travelers. 
We map  CDR coordinates to user locations with enriched contextual information.
In particular, we answer three questions:
\begin{enumerate}
    \item What are the behavioral patterns at the user locations based on longitudinal observations?
    \item Are there any common behavioral structures across the population? 
    \item Can we attach contextual information to user locations by labeling home and workplace? 
\end{enumerate}

Answering these questions enables us to understand better the home and workplace distribution, as well as the commuting flows, which are the building blocks to improve location-based services, including health, entertainment, transportation, and mobile commerce. 

The paper makes the following contributions. First, {we propose a method to infer home and workplace from CDR with improved accuracy. This technique can be applied to other longitudinal behavioral data with geo-locations (e.g., mobile phone GPS, WiFi, and blue information) to identify the home, the workplace, and the commuting flows.}

Second, we introduce the Normalized Hourly Presence to extract behavioral characteristics from CDR-based user locations, which can uncover the common behavioral patterns across the population using eigen-decomposition. We couple it with Principal Component Analysis (PCA) to show the existence of generalizable patterns. 
Understanding both the location and temporal presence pattern, which enables temporal and geographical mobile targeting~\citep{luo2014mobile}. 
        
Third, we use two datasets to validate the method and demonstrate its applicability to real-world problems. 
We use the MIT reality mining data set to illustrate that the technique accurately infers location labels such as home and workplace, improving upon the most commonly used existing methods by 79\% and 34\%, respectively. 
Our result also shows that existing methods relying on simplistic assumptions perform poorly in inferring home and workplace due to the complexities of human behavior and characteristics of mobile phone data. 
Moreover, {we test the method on CDR in a populated city in China to show the scalability of the method and its feasibility to uncover the behavioral patterns and labeling home/workplace in real-world settings.}

This paper proceeds as follows. Section~\ref{sec:lit} introduces related literature on inferring home and works locations based on CDR. Section~\ref{sec:method} presents the conceptual framework, the definition, and the calculation of the Normalized Hourly Presence and eigenlocation, and the clustering approach. Section~\ref{sec:evaluation} applies the method to the  MIT reality mining data and benchmarks against the existing approach. We present a case study with real-world CDR in a populated method in China. Section~\ref{sec:applications} discusses the managerial applications of our approach and Section~\ref{sec:conclusion} concludes the paper.

\section{Literature Review}
\label{sec:lit}

In this section, we review the literature in the location-based social network, CDR for transportation, geo-located data for inferring home and workplace. 
\subsection{Location-based social network} 
The location-based social network has been an active research field with many interesting industry applications~\citep{lee2016friend,guo2018combining}.
We have seen many successful location-based social media platforms, such as Foursquare, Facebook, Twitter, and Instagram. 
Many studies utilize information about contents, friendship networks, and mobility, facilitating companies to make better location recommendations~\citep{shi2013network,guo2018combining}. 

There has been a significant interest in location recommendation based on trajectory histories and utilize information from similar users~\citep{ghose2019mobile, zheng2010collaborative, qiu2018learning}. 
For example, \cite{ghose2019mobile} find that mobile targeting using trajectory information enables higher redemption probability and faster redemption behavior. 
\cite{shi2013network} find that weighting friends' check-ins by proximity measure can better predict a new visit to a business. 
\cite{luo2014mobile} use an experiment to show both the temporal and geographical information individual improve the performance of targeting. 
For friend recommendation, ~\cite{zheng2011recommending} focus on friend recommendation by suggesting individuals with similar interests. 
Precisely, they estimate individual interests based on sequence and visited popularity in their location histories. 
In event recommendation, \cite{macedo2015context} build an event recommendation system with data from Meetup.com, by exploiting social signals based on group memberships, location histories, and temporal signals derived from the users' time preferences. 
These studies highlight the effectiveness of using behavioral information to improve location-based services.

In this study, we aim to infer the home and workplaces from Call Detail Records, which can be used as the inputs to many location-based services, such as transportation planning. location-based targeting, offline social events recommendations, as discussed in Section~\ref{sec:applications}. 
The method can also be applied to other types of data with longitudinal behavioral information. 

\subsection{CDR for human mobility} 

{Applying CDR to transportation is a rapidly emerging research field. 
Its popularity in academia and industry results from CDR's relatively high resolution, wide availability, and high penetration rate,  comparing with traditional surveys. 
All of these benefits enable the usage of CDR in solving critical transportation problems and relieving inefficiencies in the urban environment.
Applications range from predicting OD-matrices at a macro-level to understanding travel behaviors at a micro-level \citep{leng2016urban}. 

Let us name a few examples in the literature, including travel demand management, tourist event analytics, inferring traffic mode, understanding trip purposes, and estimating land usage. 
\cite{leng2017synergistic} develop a method to make location recommendations to simultaneously improve tourist travel experiences while balancing traffic flows to reduce traffic congestion. 
\cite{wang2010transportation} and \cite{doyle2011utilising} develop a novel method to infer transportation mode based on CDR. 
Disentangling travelers with different travel modes enable transportation planners to understand current travel demand better and hence adjust traffic supply accordingly. 
\cite{leng2016analysis} develop indicators to evaluate special events for the tourism industry. 
This method enables urban planners and event organizers to assess marketing strategies, evaluate event performances, and understand tourist segmentation. 
\cite{calabrese2014} and \cite{Diao18092015} infer travel activities from CDR by integrating it with the travel diary surveys. 
Results from these studies enable urban planners to monitor traffic flow dynamically and to infer trip purposes without the need to conduct travel diary surveys daily. 
 Activity-based travel models and transportation policies can benefit from this information. 
\cite{phi2010} use CDR to infer the activity patterns of the residents and compare these patterns with the designed land use by the government. 
Understanding the discrepancies and the alignment of activities versus land-use facilitates urban planners to monitor urban dynamics better and inform future urban planning. 

}

\subsection{Geo-located data to infer home and workplace}
With the increasing penetration rate of mobile phones, passive mobile location data have the potential to be used for the inference of significant locations \citep{Ahas2010}. There is substantial research on inferring home and workplace using CDR. Most of the studies make strong assumptions to capture the corresponding behavioral patterns at home or workplace. 

The most widely-used methods of inferring home and workplace assume that home and workplace are the two places where people visit the most frequently. A couple of researchers identify home and workplace as the ones with the most significant number of presences during home and work hours~\citep{Cala2011, phi2010, jiang2013}. This assumption is limiting given the characteristics of CDR. For example, people may use landlines at home or in the workplace. Further, this assumption restricts each individual to have precisely one home and one workplace. However, not all users have at least one detectable home and workplace from CDR due to the event-driven characteristics of the data (e.g., people may not use phones at home), or they may have multiple homes and workplaces.

\cite{Kung2014, Sun2014} use dwelling time (i.e., the time staying at one location) as the feature to segment home and workplace. However, due to the event-driven and sparsity nature of CDR, the dwelling time may be over-estimated. Note that similar characteristics apply to other behavioral data collected by mobile apps or WiFi connections. 

Some studies use small-scale experiments to calibrate parameters and calculate thresholds for segmentation. \cite{Ahas2010} experimented on 14 individuals to compute the mean and the standard deviation of the earliest call to differentiate home and workplace. However, a parameter calculated from a small-scale experiment is not a representative measure to be used in other studies. 

All of these studies rely on some assumptions on the frequencies, dwell time, or earliest call in the day. At the same time, they assign a threshold to divide home and work hours, which can be arbitrary. In this study, we develop a method relying on the data-driven behavioral pattern, without assuming the presence patterns at home or define an ad-hoc threshold. Our approach also allows an estimation of the home-workplace hours and commuting time~\citep{li2012embedding,bhat2001modeling}.

\section{Methodology}
\label{sec:method}
In this section, we describe our {data-driven method} to map CDR-based user locations to the home, workplace, and {third place} based on longitudinal geographic coordinates. Two basic terms used throughout this paper are \emph{user location} and \emph{presence}. 
\begin{definition}\textbf{User location}\\
We define a user location as the weighted centroid of a cluster of cell towers that approximates the exact locations of a user. 
\end{definition}

\begin{definition}\textbf{Presence}\\
We define the {presence} as the appearance of a user at a {user location}.
\end{definition}

\subsection{Data description}
CDR records all phone users' traces with timestamps and approximate locations of the users whenever they initiate phone calls, send/receive SMS, or browse the web. This data is event-driven and, hence, does not track all the places people have visited. The raw CDR contains the encrypted user ID, the timestamp, Location Area Code (LAC), cell tower ID, and the event type. 
We map the cell towers to another database to get the coordinates, where the users are located within a radius of a half to two kilometers. 

\paragraph{Characteristics and limitations of mobile phone datA} 
We now discuss some characteristics and limitations of CDR and how our method conquers such constraints.
First,  the call record is event-driven, meaning that if there is a connection between the user and the cell tower, there will be a record; otherwise, there will be no records. 
This typical characteristic applies to most mobile phone data, most of which only contain a record when the user engages with the service.  
This characteristic makes it challenging to obtain a complete travel trajectory of the individual. 
Fortunately, this does not influence our method since we propose a normalized hourly presence to aggregate behavioral information for an extended period.

Second, CDR is collected by one of the mobile carriers, hence covers a sample of the population depending on the market share of the company. 
These characteristics and limitations apply to most mobile phone-based data, which captures a small and potentially biased sample, e.g., users from certain Apps, or a sample of users using Andriod or iOS. 
Comparatively, CDR has the highest penetration rate and captures a more representative sample. 
In our study, the data stakeholders (i.e., mobile carrier) takes up around 30\% of the market, which is a representative share of the population. 
We need to scale up the inferred home and workplace if researchers are interested in the aggregate distributions of home and workplace. 

Third, the resolution (accuracy) and privacy are paradoxical. Specifically, the spatial resolution of CDR makes it challenging to pinpoint the exact x and y coordinates of the home and work locations. 
Hence, it preserves the privacy, compared with GPS and WiFi, the spatial resolution of which is around 10 ~ 25 meters. 
Though not the exact x and y coordinates, the resolution is high enough for improving location-based services. 
Moreover, if the objective is to provide transit services or needs to combine with census data for more detailed socio-demographics, CDR can be an appropriate data source. 

\subsection{Conceptual framework}
\begin{figure}[!p]
    \centering
    \includegraphics[width=.6\linewidth]{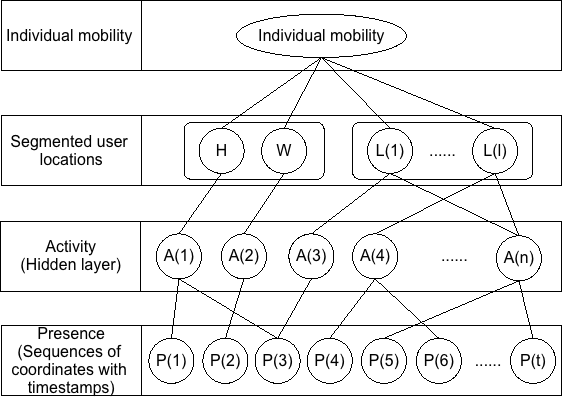}
    \caption{Conceptual Framework for Home (H), Workplace (W), and Other location (L).}
    \label{fig:concept}
\end{figure}

Figure~\ref{fig:concept} presents the conceptual framework of the data-driven method. 
For each user, we observe a sequence of coordinates with timestamps, representing the digital footprints of the user across the observational period. Each record referred to as a ``presence'', can be associated with an activity type or trip purposes. Note that there is no one-to-one correspondence between activities and user locations, meaning that people perform a particular activity at a specific user location, and people perform several activities at the same user location. Hence, the third layer, the activity layer, is unobserved and is hard to infer solely from the passive-positioning and semantic-poor CDR. People perform daily routine activities in a limited number of user locations, which can be roughly segmented into home/workplace and {third place}. Presence patterns at the user locations, shown at the bottom layer of the framework, are one of the most elementary aspects of human mobility. The objective of the {behavioral method} is to skip the activity layer and identify home/workplace from user locations based on observed longitudinal presences. CDR-based home and workplaces are different from the traditional concept of home and workplace. {CDR-based home} is the user location with home-like normalized presence patterns. Similarly, {CDR-based workplace} is the user location with weekday and weekend work-like normalized presence patterns. 

\subsection{Normalized hourly presence} 
\label{NHP}

We now introduce the proposed features to extract when, how often, and how long people appear at a particular user location. This feature captures not only the location-based presence frequency but also the temporal presence variations at the user locations on weekdays and weekends from the longitudinal records. 
To characterize the temporal presence patterns of individuals at the user locations and condense the time series data in a structured way, we propose a new feature: \textit{Normalized Hourly Presence (NHP)}. We sum the number of presences in each hour of a weekday and a weekend across the observational period to tackles the sparsity of the data. 
We aggregate the data into weekdays and weekends separately due to the diverse schedules, i.e., people's schedules are different on weekdays when they need to go to work and on weekends when they have more spare time.  We normalize the hourly presences to the percentage of presences by the total number of phone connections of the individuals. The normalization is important to resolve the high variability of the number of observations of individuals, which is another typical characteristic of this type of data collected from mobile phones.  Equation (\ref{eq:nhp1}) shows the calculation of NHP. In essence, we characterize each user location by a vector with 48 NHPs. Specifically, 

\begin{equation} \label{eq:nhp1}
    \widetilde{P}_{i,h}^{l}=\tfrac{P_{i,h}^{l}}{\sum_{l=1}^{L_{i}}P_{i,h}^{l}},
\end{equation}

\noindent where $\widetilde{P}_{i,h}^{l} $ and $ {P}_{i,h}^{l}$ are the NHPs and absolute hourly presence for individual $i$ during hour $h$ at the user location $l$. $L\in [1,48]$, representing 24 hours on weekdays and weekends. $L_{i,h}$ represents the unique set of user locations for individual $i$ during hour $h$. 

\begin{figure}[!p]
    \begin{center}\includegraphics[width=.5\linewidth]{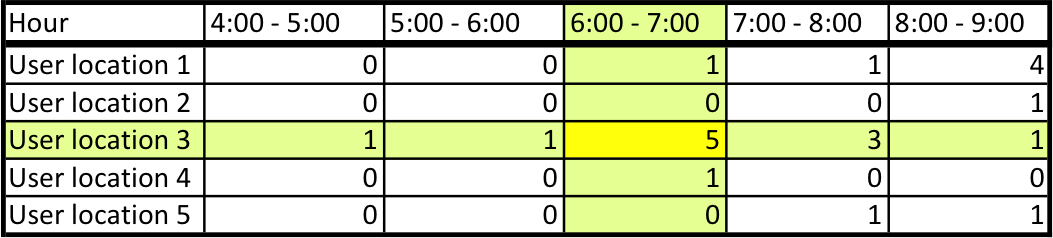}\end{center}
    \caption{Illustration on the concept of user location for one specific user.} 
    \label{nhp_eg}
\end{figure}

Figure~\ref{nhp_eg} demonstrates the computation of NHP. User A presented at 5 places for 1, 0, 5, 1 and 0 times respectively during 6:00 $-$ 7:00 on the weekday across the observational period ($L_{A}^{6} = 5$). His normalized presence at the user location $3$ during this time period can be calculated as in equation (\ref{eq:2}). Specifically, 

\begin{equation} \label{eq:2}
    \widetilde{P}_{A,6}^{3}=\tfrac{P_{A,6}^{3}}{\sum_{l=1}^{5}P_{A,6}^{l}}=\tfrac{5}{1+0+5+1+0}=\tfrac{5}{7}. 
\end{equation}

\subsection{Eigenlocations}
Due to the high regularities and generalizability of human mobility, we can extract the typical behavioral patterns and daily routines at the user locations across from the large-scale noisy user location dataset. This step is critical in revealing whether there exist common behavioral patterns across the population, and if yes, what are the presence patterns at the user locations. 

To achieve this goal, we use Principal Component Analysis (PCA) to extract the underlying structures from large-scale behavioral datasets~\citep{eagle2009, reades2009, cala2010eigen}. The resulting Eigenlocation is used to describe the common presence patterns over the weekday and the weekend across the urban-wide population. Consequently, to capture the common presence patterns across the urban-wide population, we rely on the assumption that human presence patterns at the user locations with similar functions are consistent across the sample population. Each eigenvector, named as eigenlocation, represents a typical presence pattern by explaining a portion of the variance of the behavioral presences. We rank the eigenlocations by the explained variances, which is determined by the associated eigenvalue. Projecting original presence vectors onto eigenlocations reveals common behavioral structures and reduces noisy and random behaviors. 

Next, we describe the procedure of PCA fomally.  The entire user locations set can be represented by $\Gamma _{1}, \Gamma _{2}, \Gamma _{3}, ..., \Gamma _{U}$ for a total number of $U$ user locations. $\Gamma _{U}$ is a 48 dimension vector characterized by the 48 NHPs. The average presence pattern of the user location is $\Psi =\frac{1}{U}\sum _{u=1}^{U}\phi _{u}$. For normalization, $\phi_{i}=\Gamma_i-\Psi$ is the deviation of a user location from the mean presence patterns. Let matrix $A$ be $[\phi_{1}, \phi_{3}, ..., \phi_{M}]$. The calculations are shown in Equation~(\ref{eq:3}) and Equation~(\ref{eq:4}).

\begin{equation} \label{eq:3}
    C=\frac{1}{48}\sum _{h=1}^{48}\phi _{h}\phi _{h}^{T}=AA^{T}, 
\end{equation}

\begin{equation} \label{eq:4}
    V{}'CV=\Lambda, 
\end{equation}

\noindent where $\Lambda= $diag${\left  \{ \lambda_{1}, \lambda_{2}, \lambda_{3}, ..., \lambda_{48} \right \}}$ are the eigenvalues and $V=[v_{1}, v_{2}, v_{3}, ..., v_{48}]$ is an orthogonal matrix where the $j_{th}$ column $v_{j}$ is the eigenvector correspondence to $\lambda_{j}$. 

\subsection{Clustering for user location segmentation}

We use clustering techniques to achieve the goal of extracting home and workplace from the collection of unlabeled user locations characterized by normalized hourly presences without supervision. Clustering analysis is a standard explanatory tool to discover structures and group similar objects. 
In our study, it helps to find patterns in a collection of unlabeled samples by organizing user locations that are similar in the eigenspace. 

The first task of clustering is to define the similarity measure. The similarity measure captures the closeness between presence patterns, more specifically, weekday and weekend normalized presences. We use Euclidean distance, the sum of squares of differences between normalized hourly presences with each hour equally weighted to measure similarity. The observed presence patterns reveal the function of the location for the individuals. In other words, the cluster of user locations is the group of locations that have similar meanings to users, such as home and workplace.

We apply two clustering techniques in our framework: hard clustering and soft clustering, represented by k-means clustering and Fuzzy C-means Clustering (FCM). K-means clustering assigns a single cluster label to each user location. FCM, on the contrary, attributes a vector of membership of belonging to each user location cluster (home, workplace, {third place} in our case). The method assigns each user location to the cluster with the largest ``membership,'' which we interpret as the probability of belonging to that cluster. There exist uncertainties in inferring the home and workplace due to the complexities in the data (e.g., sparsity and noise) and in human behavior (e.g., people may use landlines at home or workplaces; or may have multiple home and workplaces). 
FCM, though not as interpretable as the k-means clustering, captures the uncertainty of the inference result and allows the trade-off between certainties and inference rate for different application purposes. 

\section{Evaluation and applications}
\label{sec:evaluation}
As an evaluation, we first apply the method summarized in Algorithm~\ref{alg:home_work} to a small-scale experiment with the ground-truth to test and evaluate the proposed method through three metrics, namely accuracy, inference rate, and flexibility. 
Then, we implement the method on the real-world, large-scale dataset in a populated city in China to illustrate the practicality and scalability of our method of inferring home,  workplace, and {the third place}. 
We also show common behavioral structures, which reveal interesting and interpretable patterns. 


\subsection{Data}

    \subsubsection{Small-scale experiment}

    To show the performance of our method, we use small-scale experimental data with labeled ground truth. The Reality Mining project was conducted from 2004 to 2005 at the MIT Media Laboratory~\citep{eagle2006reality}. The Reality Mining study followed more than 100 subjects for more than a year (including students and faculty), 73 of whom are usable. The researchers tracked the subjects by mobile phones, which were pre-installed with software to record data about call logs, cell tower IDs, and phone status (idling or charging). 
The subjects labeled the records as home, workplace, {the third place}, and no-signal.

    \subsubsection{Real-world data}

    The large-scale CDR we used covers two months in a populated and fast-developing city in China. 
One of the three mobile carriers in China provides the data. We random sample of 100,000 mobile phone users and  217,753 user locations as a case study to test the scalability and the feasibility of our method. 
    
    The real-world data is much noisier than the MIT Reality Mining data. Hence, we preprocess the CDR in a Chinese city in the following way. For each user, several pass-by cell towers are clustered into one user location using the algorithm developed by~\cite{Isaacman2011}. This method first ranks the cell towers according to the total number of days that they are connected. The next step is to cluster cell towers according to Hartigan's leader algorithm with a spatial threshold of 1 km. This algorithm starts from the first cell tower in the sorted list as the center of the cluster.
The algorithm checks the subsequent cell tower to see if the record falls within the radius of one kilometer. If it does, the algorithm groups the cell tower into an existing cluster. Otherwise, it becomes a new cluster centroid. The algorithm completes when every cell tower belongs to a cluster. Readers interested in the detailed implementation of the method are encouraged to refer to ~\cite{Isaacman2011}. 
    
\subsection{Results analysis and comparisons}
\label{res}

    We compare the proposed method with the most widely-used method in the literature, the \textit{Most Frequent Appearance} method, on the MIT Reality Mining data. As stated in Section~\ref{sec:lit}, the \textit{Most Frequent Appearance} method labels the user location with the most presences during the home time (00:00 - 08:00 and 19:00 - 24:00) are home and daytime (09:00 - 18:00) are workplace respectively~\citep{Cala2011, phi2010, jiang2013}.
    Two metrics are used for performance comparisons: \textit{accuracy} and \textit{inference rate}. We use the percentage of the correct inferences as the accuracy measure. We compute the inference rate as the percentage of inferred home, workplace, and {the third place}. We show the comparisons of the results in Table~\ref{tabel:methcomp}. There exists a trade-off between the accuracy and inference rate, meaning that the larger the number of home/workplace inferred, the more likely they are incorrectly predicted. 

    The \textit{Most Frequent Appearance} method identifies one home and one workplace for every individual. The accuracies are 53\% and 62\%, respectively. K-means clustering infers 56\% home and 82\% workplace with 90\% and 75\%  accuracy. 
    FCM infer 58\% home and 84\% workplace with 88\% and 74\% accuracy. Though the inference rates are relatively low compared to the ``Most Frequent Appearance'', our method improved the accuracy considerably.

    FCM has the flexibility to trade-off accuracy and inference rate for different application purposes. To improve inference rate, we can relabel {third place} whose membership is less than the median as either home or workplace. This method can identify 78\% home and 100\% workplace, respectively, and the accuracies are 91\% and 74\%. The accuracy of the workplace stays the same compared to simple FCM, while the accuracy of home increases to 81\%. Contrarily,
    the improvements in accuracy are 79\% and 55\% for home and workplace, respectively, by relabeling the user locations with less-than-first-quartile membership to home and workplace. Relabeling home and workplaces with low membership by increasing the membership threshold helps to increase the accuracy. 
    The improvement in accuracy in workplace inference is more significant compared to that of home. It is therefore not reasonable to remove the ``home'' with lower than first quartile membership since the accuracy improvement is small compared with the decrease in inference rate. 
    Depending on the objective of the application, the policy-makers can determine the threshold for relabeling. 
    Accuracy is essential if the aim is to make personalized recommendations or interventions,  
    The inference rate might be more meaningful if the objective is to design and improve public transit systems. 

\begin{table*}[!p]
            \centering
            \caption{Methods comparison}
            \label{tabel:methcomp}
            \footnotesize
            \begin{tabular}{|l|l|l|l|l|}
                \hline
                &                {\bf Metric}      & {\bf Home} & {\bf Workplace} & {\bf Improvement (home\&workplace)} \\ \hline
                \multirow{2}{*}{{\bf ``Most Frequent Appearance''}} & {Accuracy}       & 0.53       & 0.62    
                & \multirow{2}{*}{NA}         \\ \cline{2-4} 
                & {Inference rate} & 100\%         & 100\%    &          \\ \hline
                \multirow{2}{*}{{\bf K-means clustering}}         & {Accuracy}       & 0.90       & 0.75     
                & 79\% \& 34\%       \\ \cline{2-5} 
                & {Inference rate} & 56\%         & 82\%   
                & NA           \\ \hline
                \multirow{2}{*}{{\bf FCM (Balanced)}}   & {Accuracy}       & 0.88       & 0.74     
                & 74\% \& 32\%        \\ \cline{2-5} 
                & {Inference rate} & 58\%         & 84\%   
                &   NA     \\ \hline
                \multirow{2}{*}{{\bf FCM (Prioritizing Inference Rate)}}    & {Accuracy}       & 0.91       & 0.74    & 81\% \& 32\%        \\ \cline{2-5} 
                & {Inference rate} & 78\%         & 100\%     
                & NA         \\ \hline
                \multirow{2}{*}{{\bf FCM (Prioritizing Accuracy)}} & {Accuracy}       & 0.90       
                & 0.83    & 79\% \& 55\%            \\\cline{2-5} 
                & {Inference rate} & 42\%         & 63\%    
                & NA            \\ \hline
                
            \end{tabular} 
\end{table*}

\subsection{Real-world application}

In this section, we apply our method to real-world CDR  collected in a populated city in China. 
We first analyze the pattern of eigenlocations, demonstrating that it is possible to use eigen-decomposition to extract eigenlocations as a way to reveal the characteristic presence patterns at the user locations. After that, we use the Davies-Bouldin Index as a way to determine the optimal number of user location clusters. Moreover, we analyze and interpret the patterns of the mean of each cluster. 
We examine the uncertainties of each user location cluster. Finally, we show the scalability and efficiency of the method. 

\subsubsection{Eigenlocations}

\begin{figure}[!p]
    \centering
    \includegraphics[width=\linewidth]{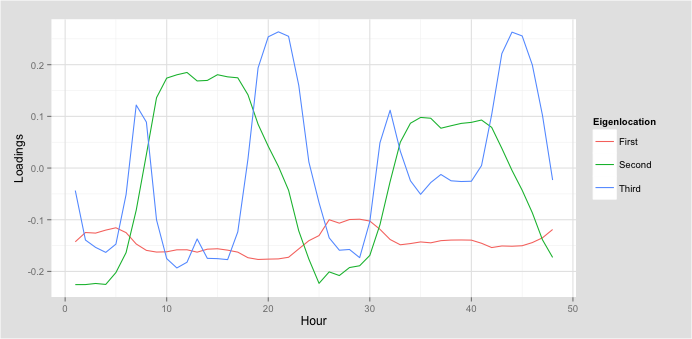}
    \caption{Top three eigenlocations. The x-axis is the hour of weekday and weekends. The y-axis correspond to the loadings of the eigenlocations at each hour of day.}
    \label{fig:eigen}
\end{figure}

Eigenlocations, each representing a typical pattern at the user locations, are shown in Figure \ref{fig:eigen}. The x-axis corresponds to the 24 hours on a weekday and 24 hours on the weekend. Moreover, the y-axis corresponds to the loadings of each hour on the eigenlocations. Substantial positive loadings indicate high presence frequency at the user locations, and negative loadings with large-magnitude indicate the low frequency at these user locations. Small magnitude, irrespective of the sign, suggests no information in the data for prediction. The red line, representing the first eigenlocation explaining 30.0\% of the variance, displays a pattern of infrequent-visiting location. The second eigenlocation (explaining 10.3\% variance), as shown in the blue line, corresponds to a daytime-active location. Large positive coefficients on this eigenlocation are an indicator of the workplace. The third eigenlocation, as illustrated in the green line, represents home-like user location where individuals mostly stay during the evenings and early mornings. 

From the figure, we find that the first eight eigenlocations are interpretable. To segment user locations, we determine the optimal number of eigenlocations be eight. This number is plausible since the non-interpretable patterns are more likely to be noise. The eight eigenlocations explain 56\% of the behavioral variances of user locations in total. We can reconstruct the presence patterns at the user locations with a  linear combination of the eight eigenlocations to reduce noise. 

\subsubsection{Optimal number of cluster}

In the inference of home and workplace, prior knowledge and hypothesis are there should exist four clusters, including one workplace, one {the third place} and two home locations, one for day-time workers and one for non-workers or short-distance commuters. 

To test the validity and stability of the optimal number of clusters, we bootstrap the Davies-Bouldin (DB) index for the different number of clusters~\citep{DB1979, Kerr2001}. DB index measures the scatter within the cluster and separation between clusters by the distances between each observation and its most similar ones. Accordingly, the lower the DB index, the better the cluster configuration. With the DB index on a different number of clusters, we can see that there is an increase in the DB index when the cluster number increases from 4 to 5, while the decrease in DB indexes is small when the number of clusters exceeds 5. This result is in line with our intuition, and the four clusters correspond to one workplace, one {third place} and two home locations, one for normal-schedule workers and one for non-workers or short-distance commuters. 


\subsubsection{Clustering results}
\begin{figure}[!p]
    \begin{center}\includegraphics[width=\linewidth]{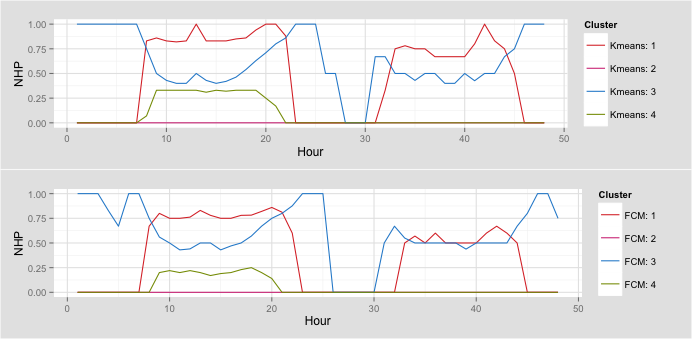}\end{center}
    \caption{Clustering results from K-means clustering and Fuzzy C-means clustering. X-axis is the hour of weekday and weekend. Y-axis is the average NHP of each cluster.}
    \label{fig:cluster}
\end{figure}

Figure~\ref{fig:cluster} presents the clustering results from k-means clustering and FCM, where the x-axis and y-axis represent 24  hours on weekdays and 24 hours on weekends, and the median NHP, respectively. Each line corresponds to the 48 median NHPs for one of the four clusters, including two types of home locations (red and blue), one workplace cluster (green), and {third place} cluster (purple). 

Overall, the results from the two clustering techniques are similar. The performances are in line with the actual home and workplace patterns in daily lives. The red line represents the home cluster for non-commuters or commuters who work near home. The percentages of presences at these locations are quite high throughout the week from 7:00 to 24:00. Note that for those who work near home, it is challenging to differentiate home and workplace due to the low spatial resolution of CDR. The blue line represents home for regular commuters, who are present at these user locations early in the morning and late at night. The presence frequencies are high before 7:00, possibly due to the automatic data fetching by some mobile apps. The green line represents the work cluster. Commuters are present at these locations more frequently from 9 am to 8 pm on weekdays. The purple line represents a type of location clustering with infrequent and irregular presences, which we label as a third place. 

\begin{figure}[!p]
    \begin{center}\includegraphics[width = \linewidth]{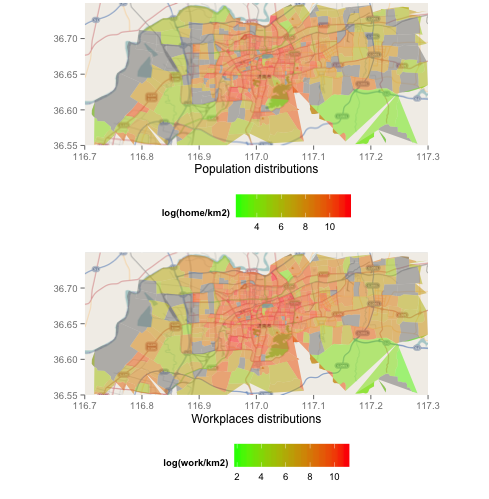}\end{center}
    \caption{Home and workplace distributions. The color corresponds to the number of home and works per km$^2$ in log-scale, identified by our method.}
    \label{hwdist}
\end{figure}

We also show the home and workplace distributions in the urban area in Figure \ref{hwdist}. We show that the population and workplace density distributions of the Traffic Analysis Zones (TAZs). We scale up the 100,000 random samples to the whole population with the values in logarithmic transformation for better visualization. The color scales locate underneath the figures. From the map, we see that home and workplace rare distributed the most densely in the city center, which is in line with reality. The density of the workplace in the eastern urban area is higher than that of the home density, where there is a high-tech district with many new employment opportunities created. 

\subsubsection{Uncertainty of the inference results}
\begin{figure}[!p]
    \centering
    \includegraphics[width=\linewidth]{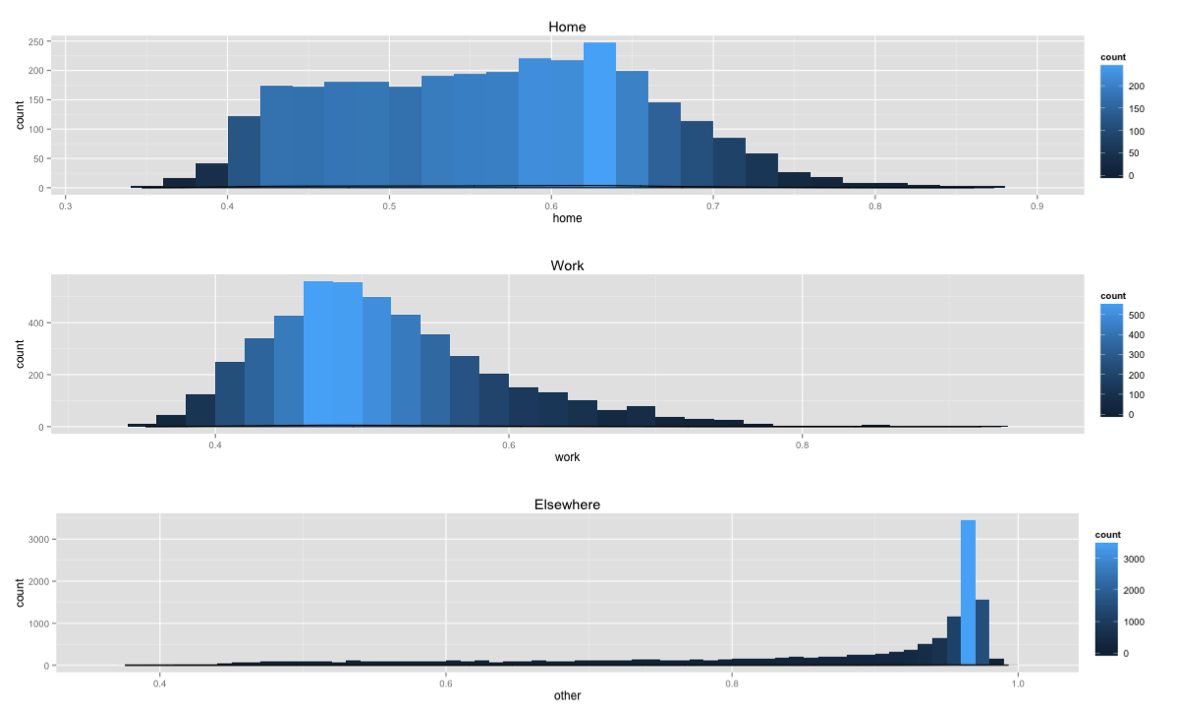}
    \caption{Membership to each cluster. The upper, middle and lower panel correspond to home, work and the third place. The x-axis and the y-axis correspond to the membership and the number of estimated home, workplace and the third place. }
    \label{fig:poss}
\end{figure}

We learn the uncertainties of the inference results from the membership assignment via FCM. The larger the membership, the more certain that the location belongs to a specific group. 
We show the membership distributions for each cluster in Figure~\ref{fig:poss}. The x-axis denotes the membership, and the y-axis represents the count of user locations in each membership range. Lighter color indicates a more substantial number of user locations. The median memberships are 0.56, 0.50, and 0.94, respectively, for the home, workplace, and the {third place}. We can see that most places belong to the {third place} with a large membership since the user locations with a small number of presences are very likely to belong to this group. 
The uncertainty for labeling the workplace is the highest since people are more active during the daytime, making the inference more difficult. 

The most significant advantage of applying FCM in this setting is the flexibility in trading-off between accuracy and inference rate using the membership as discussed above. If we increase the accepted membership threshold and hence infer fewer home/workplace, by only labeling home and workplace with membership higher-than-threshold as home/workplace. On the contrary, reducing the accepted threshold will improve the inference rate. 

\subsubsection{Computation time}
Computational complexity is an important consideration for practical applications. The computational complexity of k-means clustering is $O(ncdt)$ and that of FCM is $O(ncdt^{2})$. $n$ is the number of observations, which is the total number of user locations for all sampled mobile phone users. $d$ is the number of features, which is 48. $c$ is the pre-specified number of clusters. $t$ is the number of iterations until convergence. We test the method on 100,000 mobile phone users with 217,753 user locations; each has a two-month presence. The running time for PCA is 15 seconds. The running time for k-means clustering is approximately 6.2 seconds, and that for FCM clustering is 120.2 seconds. While k-means clustering outperforms FCM in computation time, they are both efficient, practical, and scalable in practice. 

\section{Managerial applications}
\label{sec:applications}
In this section, we discuss applications of our method in transportation planning, location-based mobile targeting, and offline social events recommendation. 

\subsection{Location-based mobile targeting}
The mobile phones serve as both a channel to distribute advertisements and an opportunity for marketing companies to target the customers based on a contextual understanding of the locations of individuals~\citep{xu2019forward,ghose2019mobile}. 
Therefore, identifying the home and workplace enables businesses to perform more effective context-based targeting by either utilizing the two types of contexts or the roads where the consumers pass by on their commute~\citep{gronroos1990relationship}. 
For example, \cite{ghose2019seizing} find that mobile coupons have differential effects on commuters and non-commuters, and thereby require different strategies at different moments.
Moreover, businesses can use outdoor broadcast mediums as well as mobile phones to achieve maximum advertisement results~\citep{lai2017improved}. 
For example, companies can identify a large group of people who reside or work in similar areas. Then, they can set up outdoor advertisements close to the train station near their workplaces, or along the highways where these people commute. Meanwhile, they can push ads to users' mobile phones when they are at home (if the consumers opted in for such an alert) or send brochures in mails to their home address. In this way, they can immerse the customers in the advertisement, which can be useful for some products. 

\subsection{Offline social events recommendation}
Offline social events can facilitate user interactions and bond local communities. 
Event recommendation is different from product or movie recommendation. 
Events are organized at a specific time during which the participants attend the event in person. 
Using behavioral presence pattern to attach contextual understanding to locations have two prominent advantages for event recommendation. 
First, people have different social roles at home and workplace; hence, the events interest them are different as they switch from a role from at work to family. 
For instance, users may be interested in professional meetups around the workplace, while family events at home. 
Second, temporal presence patterns at different user locations help to make the event recommendation more relevant for the users. 
For example, if a user works from late afternoon to midnight, a family event (e.g., run together event happening in the morning)  can be more attractive, rather than an early evening event. 

\subsection{Transportation and urban planning}
Our method has essential applications in transportation planning since home and workplace are the two most frequently visited locations. 
Our approach can be used in conjunction with census or travel surveys to help transportation planners and operators improve their services.
Let us name a few applications. 

First, identifying these places enables the public transport department to understand the travel demand better and provide services accordingly. Peak-hour commuting flows are usually the bottleneck of transportation services. Hence, accurate inference on these two places helps to improve the transit service shortage in cities. Since our method builds upon data-driven behavioral patterns, we can identify non-commuters, short, and long commuting flows. Different types of commuting flow require various transportation services. For example, for long-distance commutes, the transportation department may offer non-stop public transit to reduce the commuting time. While for regions with short-distance commuters or home-workers, rental bicycles or buses with more frequent stops are more effective. 

Second, our method can dynamically monitor urban dynamics, which can be especially helpful in developing countries where new residential buildings and job opportunities emerge more frequently. Though the urban planning department may have information about the changes, it is hard to monitor the commuting flows (origin-destination pairs) dynamically. Moreover, if a large number of people switch jobs, this dynamic monitoring can help to keep up to date with the dynamic demand. Adapting services dynamically in response to the urban dynamics help the cities to be ``smarter''.

Third, other than the commuting need, we identify the third place, which is the frequently-visited but non-home and workplace. People travel to these places in non-work hours. Knowing these places together with their homes, the transportation department can adapt the transit schedules and provide services if the current supply does not satisfy the high demand between residential/ workplace to the typical third places. Note that we may experience a seasonal change in third place. For example, during summers, people may go to the beach or swimming pools more often, while during winters, people may go skiing or to spas instead. Therefore, we may expect the third place to change with seasons, and this further highlights the advantage of our approach, which can dynamically monitor the urban dynamics without the need to collect survey data. 


\section{Conclusions}
\label{sec:conclusion}
The widely-penetrated mobile phone data provides longitudinal records for tracking human mobility. The low spatial resolution and sparse sampling characteristics make it challenging to utilize this promising data set in electronic commerce and urban planning. Home and workplace---origins and destinations of commuting and other trips--- are the most crucial user locations and are the building blocks of location-based services and transportation research. However, the existing method relies on strong assumptions, making it inaccurate in inferring home and workplace. 

We provide a novel method to infer the two critical locations, home, and workplace, using passively-collected mobile phone data. Our method is based on behavioral patterns at locations and performs much better than existing methods. We apply our approach to a real-world dataset in a populated city in China to show the scalability of the technique. 
We take advantage of the behavioral information in the mobile phone dataset. This opportunistic dataset exists in almost every country in the world so that the method can be widely applicable. Moreover, our approach is generalizable to any other geo-located datasets, such as WiFi, GPS, or social platforms. 

We acknowledge that there are some limitations to our data-driven method. 
The home and workplaces of mobile phone users with flexible work locations (e.g.,  shippers and drivers) or workers with uncommon work schedules (e.g., night shifters) are undetectable or prone to be misidentified. 
This misspecification can be driven from the unusual pattern, which exhibits only in a small portion of the population, and is thus hard to observe from the eigenlocation reconstruction. Another limitation is that phone usage patterns can affect the inference result. For example, if individuals use landlines instead of mobile phones at home, it is hard to estimate a home or workplace for these individuals due to the unobservable presence patterns solely based on CDR.

Though behavioral data helps to improve services for consumers~\citep{hardjono2019trusted},  some malicious companies might abuse data, which poses a high risk to society. 
Therefore, we need legislation and regulations to reduce risks, such as the Consumer Privacy Bill of Rights proposed by the Obama administration and the EU's data protection directives. 
Data companies should respect consumers' benefits and preferences on Data~\citep{stopczynski20196}.
Meanwhile, there is a great need for a secure platform to take advantage of behavioral data while preserving data security. 
MIT Trust::DATA Consortium and OPAL platform allow researchers and practitioners to manage access to the data more securely, efficiently, and equitably, while more importantly, protecting personal data from incursion and corruption~\citep{penland2016towards,hardjono2019trusted}. 

This work points out several future directions. The proposed method can extend to infer not only home and workplace but also other user locations, such as late-night locations or weekend locations. The clusters can be improved to label more types of user locations for different applications as future research. 
Besides, our method can extend to estimate commuting characteristics (e.g., commuting distances, departure, and arrival times) for better transit design. 
Future studies can extend our approach to combine CDR with other geographical data sources, such as land use data, and Points of Interest, to infer trip purposes and activity types. With more data available,  such as online-social networks check-ins (Flickr, Twitter), and bank transactions, future studies can enrich the method to provide a deeper understanding of human behaviors to enhances managerial information systems.


%
%
%




\bibliographystyle{informs2014}
\bibliography{ISRE-template} 

\begin{thebibliography}{49}
\providecommand{\natexlab}[1]{#1}
\providecommand{\url}[1]{\texttt{#1}}
\providecommand{\urlprefix}{URL }

\bibitem[{Ahas et~al.(2010)Ahas, Silm, J{\"a}rv, Saluveer, \protect\BIBand{}
  Tiru}]{Ahas2010}
Ahas R, Silm S, J{\"a}rv O, Saluveer E, Tiru M (2010) Using mobile positioning
  data to model locations meaningful to users of mobile phones. \emph{Journal
  of Urban Technology} 17(1):3--27.

\bibitem[{Andrews et~al.(2016)Andrews, Luo, Fang, \protect\BIBand{}
  Ghose}]{andrews2016mobile}
Andrews M, Luo X, Fang Z, Ghose A (2016) Mobile ad effectiveness:
  Hyper-contextual targeting with crowdedness. \emph{Marketing Science}
  35(2):218--233.

\bibitem[{Banerjee \protect\BIBand{} Dholakia(2008)}]{banerjee2008mobile}
Banerjee SS, Dholakia RR (2008) Mobile advertising: Does location based
  advertising work? \emph{International Journal of Mobile Marketing} .

\bibitem[{Bhat(2001)}]{bhat2001modeling}
Bhat C (2001) Modeling the commute activity-travel pattern of workers:
  formulation and empirical analysis. \emph{Transportation Science}
  35(1):61--79.

\bibitem[{Blondel et~al.(2015)Blondel, Decuyper, \protect\BIBand{}
  Krings}]{blondel2015}
Blondel VD, Decuyper A, Krings G (2015) A survey of results on mobile phone
  datasets analysis. \emph{EPJ Data Science} 4(1):1--55.

\bibitem[{Calabrese et~al.(2011)Calabrese, Colonna, Lovisolo, Parata,
  \protect\BIBand{} Ratti}]{Cala2011}
Calabrese F, Colonna M, Lovisolo P, Parata D, Ratti C (2011) Real-time urban
  monitoring using cell phones: A case study in rome. \emph{Intelligent
  Transportation Systems, IEEE Transactions on} 12(1):141--151.

\bibitem[{Calabrese et~al.(2014)Calabrese, Ferrari, \protect\BIBand{}
  Blondel}]{calabrese2014}
Calabrese F, Ferrari L, Blondel VD (2014) Urban sensing using mobile phone
  network data: A survey of research. \emph{ACM Comput. Surv.}
  47(2):25:1--25:20, ISSN 0360-0300,
  \urlprefix\url{http://dx.doi.org/10.1145/2655691}.

\bibitem[{Calabrese et~al.(2010)Calabrese, Reades, \protect\BIBand{}
  Ratti}]{cala2010eigen}
Calabrese F, Reades J, Ratti C (2010) Eigenplaces: segmenting space through
  digital signatures. \emph{Pervasive Computing, IEEE} 9(1):78--84.

\bibitem[{Davies \protect\BIBand{} Bouldin(1979)}]{DB1979}
Davies DL, Bouldin DW (1979) A cluster separation measure. \emph{Pattern
  Analysis and Machine Intelligence, IEEE Transactions on} (2):224--227.

\bibitem[{Diao et~al.(2015)Diao, Zhu, Ferreira, \protect\BIBand{}
  Ratti}]{Diao18092015}
Diao M, Zhu Y, Ferreira J, Ratti C (2015) Inferring individual daily activities
  from mobile phone traces: A boston example. \emph{Environment and Planning B:
  Planning and Design}
  \urlprefix\url{http://dx.doi.org/10.1177/0265813515600896}.

\bibitem[{Doyle et~al.(2011)Doyle, Hung, Kelly, McLoone, \protect\BIBand{}
  Farrell}]{doyle2011utilising}
Doyle J, Hung P, Kelly D, McLoone SF, Farrell R (2011) Utilising mobile phone
  billing records for travel mode discovery .

\bibitem[{Eagle \protect\BIBand{} Pentland(2006)}]{eagle2006reality}
Eagle N, Pentland A (2006) Reality mining: sensing complex social systems.
  \emph{Personal and ubiquitous computing} 10(4):255--268.

\bibitem[{Eagle \protect\BIBand{} Pentland(2009)}]{eagle2009}
Eagle N, Pentland AS (2009) Eigenbehaviors: Identifying structure in routine.
  \emph{Behavioral Ecology and Sociobiology} 63(7):1057--1066.

\bibitem[{Ghose et~al.(2019{\natexlab{a}})Ghose, Kwon, Lee, \protect\BIBand{}
  Oh}]{ghose2019seizing}
Ghose A, Kwon HE, Lee D, Oh W (2019{\natexlab{a}}) Seizing the commuting
  moment: Contextual targeting based on mobile transportation apps.
  \emph{Information Systems Research} 30(1):154--174.

\bibitem[{Ghose et~al.(2019{\natexlab{b}})Ghose, Li, \protect\BIBand{}
  Liu}]{ghose2019mobile}
Ghose A, Li B, Liu S (2019{\natexlab{b}}) Mobile targeting using customer
  trajectory patterns. \emph{Management Science} 65(11):5027--5049.

\bibitem[{Gonzalez et~al.(2008)Gonzalez, Hidalgo, \protect\BIBand{}
  Barabasi}]{gonzalez2008understanding}
Gonzalez MC, Hidalgo CA, Barabasi AL (2008) Understanding individual human
  mobility patterns. \emph{nature} 453(7196):779.

\bibitem[{Grauwin et~al.(2015)Grauwin, Sobolevsky, Moritz, G{\'o}dor,
  \protect\BIBand{} Ratti}]{gr2015}
Grauwin S, Sobolevsky S, Moritz S, G{\'o}dor I, Ratti C (2015) Towards a
  comparative science of cities: using mobile traffic records in new york,
  london, and hong kong. \emph{Computational approaches for urban
  environments}, 363--387 (Springer).

\bibitem[{Gronroos(1990)}]{gronroos1990relationship}
Gronroos C (1990) Relationship approach to marketing in service contexts: The
  marketing and organizational behavior interface. \emph{Journal of business
  research} 20(1):3--11.

\bibitem[{Guo et~al.(2018)Guo, Zhang, Fan, \protect\BIBand{}
  Li}]{guo2018combining}
Guo J, Zhang W, Fan W, Li W (2018) Combining geographical and social influences
  with deep learning for personalized point-of-interest recommendation.
  \emph{Journal of Management Information Systems} 35(4):1121--1153.

\bibitem[{Hardjono et~al.(2019)Hardjono, Shrier, \protect\BIBand{}
  Pentland}]{hardjono2019trusted}
Hardjono T, Shrier DL, Pentland A (2019) \emph{Trusted Data: A New Framework
  for Identity and Data Sharing} (MIT Connection Science \& Engineering).

\bibitem[{Hasija et~al.(2020)Hasija, Shen, \protect\BIBand{}
  Teo}]{hasija2020smart}
Hasija S, Shen ZJM, Teo CP (2020) Smart city operations: Modeling challenges
  and opportunities. \emph{Manufacturing \& Service Operations Management} .

\bibitem[{Hong \protect\BIBand{} Tam(2006)}]{hong2006understanding}
Hong SJ, Tam KY (2006) Understanding the adoption of multipurpose information
  appliances: The case of mobile data services. \emph{Information systems
  research} 17(2):162--179.

\bibitem[{Isaacman et~al.(2011)Isaacman, Becker, C{\'a}ceres, Kobourov,
  Martonosi, Rowland, \protect\BIBand{} Varshavsky}]{Isaacman2011}
Isaacman S, Becker R, C{\'a}ceres R, Kobourov S, Martonosi M, Rowland J,
  Varshavsky A (2011) Identifying important places in people’s lives from
  cellular network data. \emph{Pervasive computing}, 133--151 (Springer).

\bibitem[{Jiang et~al.(2013)Jiang, Fiore, Yang, Ferreira~Jr, Frazzoli,
  \protect\BIBand{} Gonz{\'a}lez}]{jiang2013}
Jiang S, Fiore GA, Yang Y, Ferreira~Jr J, Frazzoli E, Gonz{\'a}lez MC (2013) A
  review of urban computing for mobile phone traces: current methods,
  challenges and opportunities. \emph{Proceedings of the 2nd ACM SIGKDD
  International Workshop on Urban Computing}, 2 (ACM).

\bibitem[{Kerr \protect\BIBand{} Churchill(2001)}]{Kerr2001}
Kerr MK, Churchill GA (2001) Bootstrapping cluster analysis: assessing the
  reliability of conclusions from microarray experiments. \emph{Proceedings of
  the National Academy of Sciences} 98(16):8961--8965.

\bibitem[{Kung et~al.(2014)Kung, Greco, Sobolevsky, \protect\BIBand{}
  Ratti}]{Kung2014}
Kung KS, Greco K, Sobolevsky S, Ratti C (2014) Exploring universal patterns in
  human home-work commuting from mobile phone data .

\bibitem[{Lai et~al.(2017)Lai, Cheng, \protect\BIBand{}
  Lansley}]{lai2017improved}
Lai J, Cheng T, Lansley G (2017) Improved targeted outdoor advertising based on
  geotagged social media data. \emph{Annals of GIS} 23(4):237--250.

\bibitem[{Lee et~al.(2016)Lee, Qiu, \protect\BIBand{} Whinston}]{lee2016friend}
Lee GM, Qiu L, Whinston AB (2016) A friend like me: Modeling network formation
  in a location-based social network. \emph{Journal of Management Information
  Systems} 33(4):1008--1033.

\bibitem[{Leng et~al.(2016{\natexlab{a}})Leng, Noriega, Pentland, Winder, Lutz,
  \protect\BIBand{} Alonso}]{leng2016analysis}
Leng Y, Noriega A, Pentland A, Winder I, Lutz N, Alonso L (2016{\natexlab{a}})
  Analysis of tourism dynamics and special events through mobile phone
  metadata. \emph{arXiv preprint arXiv:1610.08342} .

\bibitem[{Leng et~al.(2017)Leng, Rudolph, Zhao, \protect\BIBand{}
  Koutsopolous}]{leng2017synergistic}
Leng Y, Rudolph L, Zhao J, Koutsopolous HN (2017) Synergistic data-driven
  travel demand management based on phone records .

\bibitem[{Leng et~al.(2016{\natexlab{b}})}]{leng2016urban}
Leng Y, et~al. (2016{\natexlab{b}}) \emph{Urban computing using call detail
  records: mobility pattern mining, next-location prediction and location
  recommendation}. Ph.D. thesis, Massachusetts Institute of Technology.

\bibitem[{Li et~al.(2012)Li, Tirachini, \protect\BIBand{}
  Hensher}]{li2012embedding}
Li Z, Tirachini A, Hensher DA (2012) Embedding risk attitudes in a scheduling
  model: application to the study of commuting departure time.
  \emph{Transportation science} 46(2):170--188.

\bibitem[{Lu et~al.(2013)Lu, Wetter, Bharti, Tatem, \protect\BIBand{}
  Bengtsson}]{lu2013approaching}
Lu X, Wetter E, Bharti N, Tatem AJ, Bengtsson L (2013) Approaching the limit of
  predictability in human mobility. \emph{Scientific reports} 3:srep02923.

\bibitem[{Luo et~al.(2014)Luo, Andrews, Fang, \protect\BIBand{}
  Phang}]{luo2014mobile}
Luo X, Andrews M, Fang Z, Phang CW (2014) Mobile targeting. \emph{Management
  Science} 60(7):1738--1756.

\bibitem[{Macedo et~al.(2015)Macedo, Marinho, \protect\BIBand{}
  Santos}]{macedo2015context}
Macedo AQ, Marinho LB, Santos RL (2015) Context-aware event recommendation in
  event-based social networks. \emph{Proceedings of the 9th ACM Conference on
  Recommender Systems}, 123--130.

\bibitem[{Phithakkitnukoon et~al.(2010)Phithakkitnukoon, Horanont, Di~Lorenzo,
  Shibasaki, \protect\BIBand{} Ratti}]{phi2010}
Phithakkitnukoon S, Horanont T, Di~Lorenzo G, Shibasaki R, Ratti C (2010)
  Activity-aware map: Identifying human daily activity pattern using mobile
  phone data. \emph{Human Behavior Understanding}, 14--25 (Springer).

\bibitem[{Qiu et~al.(2018)Qiu, Shi, \protect\BIBand{}
  Whinston}]{qiu2018learning}
Qiu L, Shi Z, Whinston AB (2018) Learning from your friends’ check-ins: An
  empirical study of location-based social networks. \emph{Information Systems
  Research} 29(4):1044--1061.

\bibitem[{Reades et~al.(2009)Reades, Calabrese, \protect\BIBand{}
  Ratti}]{reades2009}
Reades J, Calabrese F, Ratti C (2009) Eigenplaces: analysing cities using the
  space-time structure of the mobile phone network. \emph{Environment and
  Planning B: Planning and Design} 36(5):824--836.

\bibitem[{Schneider et~al.(2013)Schneider, Belik, Couronn{\'e}, Smoreda,
  \protect\BIBand{} Gonz{\'a}lez}]{schneider2013unravelling}
Schneider CM, Belik V, Couronn{\'e} T, Smoreda Z, Gonz{\'a}lez MC (2013)
  Unravelling daily human mobility motifs. \emph{Journal of The Royal Society
  Interface} 10(84):20130246.

\bibitem[{Shi \protect\BIBand{} Whinston(2013)}]{shi2013network}
Shi Z, Whinston AB (2013) Network structure and observational learning:
  Evidence from a location-based social network. \emph{Journal of Management
  Information Systems} 30(2):185--212.

\bibitem[{Shrier(2016)}]{penland2016towards}
Shrier D (2016) Towards an internet of trusted data: A new framework for
  identity and data sharing.

\bibitem[{Stopczynski et~al.(2019)Stopczynski, Sweatt, Hardjono,
  \protect\BIBand{} Pentland}]{stopczynski20196}
Stopczynski A, Sweatt B, Hardjono T, Pentland A (2019) 6 the new deal on data.
  \emph{Trusted Data: A New Framework for Identity and Data Sharing} 129.

\bibitem[{Sun et~al.(2014)Sun, Yao, Wang, Qiao, \protect\BIBand{}
  Rong}]{Sun2014}
Sun L, Yao L, Wang S, Qiao J, Rong J (2014) Properties analysis on travel
  intensity of land use patterns. \emph{Mathematical Problems in Engineering}
  2014.

\bibitem[{Toole et~al.(2014)Toole, Colak, Alhasoun, Evsukoff, \protect\BIBand{}
  Gonzalez}]{toole2014path}
Toole JL, Colak S, Alhasoun F, Evsukoff A, Gonzalez MC (2014) The path most
  travelled: mining road usage patterns from massive call data. \emph{arXiv
  preprint arXiv:1403.0636} .

\bibitem[{Wang et~al.(2010)Wang, Calabrese, Di~Lorenzo, \protect\BIBand{}
  Ratti}]{wang2010transportation}
Wang H, Calabrese F, Di~Lorenzo G, Ratti C (2010) Transportation mode inference
  from anonymized and aggregated mobile phone call detail records. \emph{13th
  international IEEE conference on intelligent transportation systems},
  318--323 (IEEE).

\bibitem[{Xu et~al.(2019)Xu, Duan, Hu, Cheng, \protect\BIBand{}
  Zhu}]{xu2019forward}
Xu L, Duan JA, Hu YJ, Cheng Y, Zhu Y (2019) Forward-looking behavior in mobile
  data consumption and targeted promotion design: A dynamic structural model.
  \emph{Information Systems Research} 30(2):616--635.

\bibitem[{Zhang \protect\BIBand{} Katona(2012)}]{zhang2012contextual}
Zhang K, Katona Z (2012) Contextual advertising. \emph{Marketing Science}
  31(6):980--994.

\bibitem[{Zheng et~al.(2010)Zheng, Cao, Zheng, Xie, \protect\BIBand{}
  Yang}]{zheng2010collaborative}
Zheng VW, Cao B, Zheng Y, Xie X, Yang Q (2010) Collaborative filtering meets
  mobile recommendation: A user-centered approach. \emph{Twenty-Fourth AAAI
  Conference on Artificial Intelligence}.

\bibitem[{Zheng et~al.(2011)Zheng, Zhang, Ma, Xie, \protect\BIBand{}
  Ma}]{zheng2011recommending}
Zheng Y, Zhang L, Ma Z, Xie X, Ma WY (2011) Recommending friends and locations
  based on individual location history. \emph{ACM Transactions on the Web
  (TWEB)} 5(1):1--44.

\end{thebibliography}

\end{document}